# Pre-heating effect on the catalytic growth of partially filled carbon nanotubes by chemical vapor deposition

Joydip Sengupta[1] and Chacko Jacob[*]

[1]Materials Science Centre, Indian Institute of Technology, Kharagpur, India 721302.

E-mail: joydipdhruba@gmail.com

[*]Corresponding author: Materials Science Centre, Indian Institute of Technology, Kharagpur, India 721302. E-mail: cxj14_holiday@yahoo.com Tel: +913222-283964; Fax: +913222 255303

**Abstract**

The surface reconstruction of the Fe catalyst films due to high temperature processing in hydrogen prior to nanotube nucleation and its effect on the growth morphologies of partially filled carbon nanotubes (CNTs) synthesized using atmospheric pressure chemical vapor deposition (APCVD) of propane was investigated. Results show that pre-heating of the catalyst film deeply influences the particle size distribution, which governs the growth morphologies of the corresponding CNTs. The distribution of the catalyst particles over the Si substrate was analyzed before and after the heat treatment by atomic force microscopy (AFM) which reveals that heat treatment causes clusters of catalyst to coalesce and form macroscopic islands. The X-ray diffraction (XRD) pattern of the grown material indicates that they are graphitic in nature. Scanning electron microscopy (SEM) analysis suggested that the growth density strongly depends on the pre heat treatment of the Fe catalyst film. Multiwalled CNTs with partial catalyst filling were observed via high-resolution transmission electron microscopy (HRTEM) measurements.

The degree of graphitization of the CNTs also depends on the pre heating as demonstrated by Raman analysis. A simple model for the growth of partially catalyst filled nanotubes is proposed.

## 1. Introduction

Carbon nanotubes (CNTs) have attracted wide attention both in the research and industrial communities due to their unique structure, novel properties and potential applications[1–5], since their discovery in 1991 by Iijima[6]. The synthesis of the CNTs can be grouped into the following categories[7-11]: arc discharge, chemical vapor deposition (CVD), plasma method, laser ablation, etc. Among different methods, CVD method is simple and easy to implement, and has been widely used because of its potential advantage to produce a large amount of CNTs growing directly on a desired substrate with high purity, large yield and controlled alignment, whereas the nanotubes must be collected separately in the other growth techniques.

The general catalytic growth process of the CNTs by CVD is based on the following mechanism proposed by Baker et al.[12] known as the vapor–liquid–solid (VLS) mechanism. In this mechanism, the liquid catalytic particles at high temperature absorb carbon atoms from the vapor to form a metal-carbon solid state solution. When this solution becomes supersaturated, carbon precipitates at the surface of the particle in its stable form and lead to the formation of a carbon tube structure. The metal clusters acting as a catalyst for CNT growth can be produced by different methods, e.g. by vapor or sputter deposition of a thin metallic layer on the substrate, by saturation of porous materials by metals, by deposition of solutions of substances containing catalyzing metals or by introduction of organometallic substances into the reactor[13-16]. Prior to CNT synthesis, high temperature hydrogen treatment of the catalyst is an important step in order to produce contamination free catalyst and for the removal of the oxides that may exist over the catalyst surface[17]. During this procedure, heating above a certain temperature causes catalyst clusters to coalesce and form macroscopic islands. This process is based on cluster diffusion and depends on their density and surface diffusion constant, at a given substrate temperature. Cluster diffusion terminates when the island shape is of minimum energy for the specific annealing conditions[18]. These clusters act as a catalyst surface and the cluster size plays a critical role in CNT growth. So a detailed comparative study on the effects of pre-heating on carbon nanotube growth, morphology and microstructure is of importance to achieve a controllable growth of CNTs.

Another interesting aspect of CNT is the cavity, which can be used to incorporate metal clusters in order to generate novel nanostructured materials with new electronic or magnetic properties as a consequence of large surface-to-volume ratio of the confined materials and the interaction between the confined materials and the inner walls of nanotube[19-21]. Apart from the geometrical advantage of a cylindrical shaped nanostructure design, the carbon shells provide an effective protection against the oxidation of the confined metals and ensure their long-term stability at the core. Different filling methods in CNTs including capillary incursion[22,23] chemical method[24], arc-discharge[25,26] and chemical vapor deposition (CVD) [27,28] have already been reported. But the process is costly since it usually requires a two-stage system and rigorous control of parameters such as decomposition temperature, heating rate, precursor ratio, etc. The synthesis of filled carbon nanotube by a simple one stage process still remains a major challenge.

Here we report a systematic study of the effect of surface reconstruction of Fe catalyst films during high temperature processing in hydrogen atmosphere on the growth of partially filled carbon nanotubes (CNTs) synthesized using the atmospheric pressure chemical vapor deposition (APCVD) technique. The effect of thermal treatment on the catalyst was studied using atomic force microscopy (AFM). The morphology, internal structure and degree of graphitization of CNTs were investigated using scanning electron microscopy (SEM), high resolution transmission electron microscopy (HRTEM), X-ray diffraction (XRD) and Raman spectroscopy.

## 2. Experimental details

Atmospheric pressure chemical vapor deposition (APCVD) of CNTs was carried out by catalytic decomposition of propane on Si(111) substrates with a catalyst overlayer in a hot-wall horizontal CVD reactor using a resistance-heated furnace (ELECTROHEAT EN345T). The Si(111) substrates were ultrasonically cleaned with acetone and deionised water prior to catalyst film deposition. A thin film of iron was deposited on the substrate by a vacuum system (Hind Hivac: Model 12A4D). The substrates were then loaded into a quartz tube furnace, pumped down to $10^{-2}$ Torr and backfilled with flowing argon. When the furnace temperature stabilized at the desired temperature (850°C, 900°C and 1000°C respectively), the samples were annealed in hydrogen atmosphere for 10min. Finally, the reactor temperature was set to 850°C and the hydrogen was turned off. Thereafter propane was introduced into the gas stream at a flow rate of 200 sccm, for 1 h for CNT synthesis. The CNTs synthesized with pre heating at 850°C, 900°C and 1000°C were assigned the name a-CNT, b-CNT and c-CNT respectively.

A Nanonics Multiview 1000™ system in AFM mode was used to image the surface morphology of the Fe layer before and after the heat treatment with a quartz optical fiber tip using tapping mode. Samples were also characterized by an X-Ray diffractometer (Philips PW1729) using Cu K$\alpha$ radiation ($\lambda$=1.54059 Å) and $\theta$-$2\theta$ geometry to analyze the crystallinity and phases of grown species. Raman measurements were carried out with a RENISHAW RM1000B LRM at room temperature in the backscattering geometry using a 514.5 nm air-cooled $Ar^+$ laser as an excitation source for compositional analysis. SEM (VEGA TESCAN) and HRTEM (JEOL JEM 2100) equipped with an EDX analyzer (OXFORD Instruments) were employed for examination of the morphology and microstructure of the CNTs. The sample preparation for HRTEM study was done by scraping the nanotubes from Si substrates and dispersing them ultrasonically in alcohol and then transferred to the carbon coated copper grids.

## 3. Results and discussion

**Figure 1a** shows the AFM image of the as deposited catalyst film over the Si (111) substrate. The AFM image reveals that the initial film consists of Fe clusters instead of a continuous layer. **Figures 1b-d** show the tapping mode AFM images of the surfaces of Fe films after being annealed at 850°C, 900°C and 1000°C in hydrogen for 10 min. This heat treatment results in the formation of Fe islands as confirmed by the AFM images. However, Fe particles are not uniformly distributed on the substrate and their sizes range from about tens of nanometers to hundreds of nanometers. It is clear that the size and density distribution of Fe nanoparticles change greatly after annealing at 1000°C compared to the heat treatment at 850°C and 900°C.

In order to precisely investigate the effect of annealing temperatures on the reconstruction of Fe films, we have statistically analyzed the size distribution of the Fe particles obtained after thermal treatment in hydrogen at different temperatures (**Figure 2a-c**). Fe nanoparticles of size ranging from about 3 nm to 170 nm have multi distribution peaks after thermal treatment in hydrogen at 850°C. With increased heating temperature from 850°C to 900°C, the distribution range of the particle sizes decrease. After annealing at 1000°C the distribution range of particle sizes increases again and the particles have a bimodal distribution i.e. below and above 100 nm. The AFM images and statistical results together demonstrate that the Fe particles are refined and there is a small increase in uniformity with increasing annealing temperature from 850°C to 900°C. However, with the heating at 1000°C, large particles formed and the uniformity decreases again.

Scanning Electron Microscopy (SEM) was employed for the analysis of the morphology and density of CNTs. **Figures 3a-f** are SEM micrographs showing the surface morphology of the a-CNT, b-CNT and c-CNT. High-aspect ratio nanostructures are observed on the Fe film surfaces. In particular, the growth is very sparse in case of c-CNT (**Figure 3c**); the higher magnified image of c-CNT (**Figure 3f**) shows that the nanotubes grow only over the smaller clusters whereas the large clusters are ineffective for the nanotube growth. However for the a-CNT and b-CNT films, better results were achieved with a high density of nanotubes having randomly oriented spaghetti-like morphology. In many cases, small bright catalyst particles were detected at the tip of the CNTs. This suggests that the tip growth mechanism is likely to be responsible for the nanotube synthesis under the present conditions.

The diffusion model can be used to explain the growth of CNTs on catalyst clusters[29]. According to the diffusion model, CNTs grow by diffusion driven precipitation of carbon from the supersaturated catalyst particles. If the cluster size is larger than the diffusion length of the carbon atoms then CNTs cannot grow. In our case, at 1000 °C, the catalyst film mostly forms large clusters whereas at 850°C and 900°C, it produces relatively smaller clusters (**Figures 1b-d and 2a-c**). In case of c-CNT the size of most of the clusters is much larger than the diffusion length; therefore, carbon atoms supplied by decomposition of propane cannot diffuse into the catalyst clusters to form CNTs (**Figures 3c and 3f**). Only small amounts of CNTs grow over the smaller clusters. For a-CNT and b-CNT, most of the catalyst clusters are smaller than the diffusion length. Therefore, these clusters can effectively aid CNT growth resulting in a high density of CNTs.

High-resolution transmission electron microscopy (HRTEM) was used to characterize the growth morphology and internal structure of the nanotubes. **Figures 4a-c** show high-resolution transmission electron microscope (HRTEM) images of the a-CNT, b-CNT and c-CNT, respectively. The lattice images (**Figures 4d-f**) elucidate the effects of preheating on the degree of graphitization of the a-CNT, b-CNT and c-CNT, respectively. The CNTs exhibit a multiwalled structure for all tubes grown under the three different conditions and their diameters are below 100 nm. A catalytic nanoparticle of the size nearly 40 nm is encapsulated at the top of the a-CNT implying tip growth mechanism. A few elongated particles can be observed imbedded in the core of the nanotubes, which demonstrates the hollow nature of the nanotubes deposited in this study. Chemical composition analysis (EDX) confirms (not shown here) that the elongated particles inside the tubes are of iron.

XRD measurements were carried out to examine the structure of the CNTs and the resulting θ-2θ scan is shown in **Figure 5**. The peak at 26.2° is present in all the samples; this is the characteristic graphitic peak arising due to the presence of multiwall carbon nanotubes (MWNTs) in the sample. The peak near 43.7° is attributed to the (101) plane of the nanotube and the peak at 44.7° originates from the Fe. Intensities of all the CNT related peaks are nearly equal in case of a-CNT and b-CNT but significantly change in the c-CNT sample. The change in intensities of the XRD peaks can be explained from SEM observations. The XRD peak intensity depends upon number of CNTs present in the sample. As the growth density of c-CNT is the lowest so the CNT peaks corresponding to c-CNT have lower intensity compared to other samples. The peak at 28.5°, however, does not originate from the CNTs and is attributed to (111) plane of the Si substrate.

Raman spectroscopy provides more details of the quality and structure of the materials produced. **Figure 6** shows the room temperature Raman spectra of the MWNT material at a laser excitation wavelength of 514.5 nm. The two main peaks are the D and G bands.

The G-band represents the high frequency $E_{2g}$ Raman scattering mode of $sp^2$-hybridized carbon material i.e. the stretching mode of the C-C bond in the graphite plane and demonstrates the presence of crystalline graphitic carbon. In all cases it appears near 1575 cm$^{-1}$, which is very close to the value observed by Hiura et al.[30] for carbon nanotubes and nanoparticles.

The D-band, at around 1348 cm$^{-1}$, originates due to the presence of lattice defects in the graphite sheets that make up carbon nanotubes. The position of the D band for MWNT can be expressed as $\omega_D = 1285 + 26.5 E_{laser}$ [31]. In our case $E_{laser}$ = 2.41eV resulting in $\omega_D \approx 1349$ cm$^{-1}$, which is in agreement with our observed peak position. However, a recent Raman analysis of multiwalled nanotubes (MWNT) suggests that the D band is an intrinsic feature of the Raman spectrum of MWNTs, and they are not necessarily an indication of a disordered wall structure[32].

The ratio of the intensity of the D band and G band ($I_D/I_G$) in the Raman spectra is used extensively as a measure of the graphitization of a sample, and also used to analyze the quality of the tubes produced e.g. a smaller $I_D/I_G$ ratio corresponds to fewer defects[32]. The intensity ratio of the G band with respect to the D band varies in the three samples. The values of ($I_D/I_G$) for the a-CNT, b-CNT and c-CNT are 0.23, 0.18 and 0.20 respectively. Although the shape and positions of peaks in the spectra are almost similar, the variation in the intensity ratio of the D to G peaks (R = $I_D/I_G$) are indicative of differences in the degree of graphitization. The decrease in the relative intensity of the disordered mode for b-CNT and c-CNT samples can be attributed to the decreased number of structural defects.

In the considerable reports regarding filled CNT synthesis[33-37], the procedure followed to fill the carbon nanotubes using CVD is the floating catalyst method whereas we used fixed catalyst for nanotube growth. Here we present a plausible mechanism for the partially filled CNT growth that agrees with our experimental results.

On the basis of the aforementioned experimental results and analysis we can conclude that the growth is primarily governed by the tip growth mechanism and we suggest the following model to explain the growth mechanism of the nanotubes having a tubular structure with partial Fe filling, as schematically shown in **Figure 7**. The growth process starts with the diffusion of carbon into the metallic iron nanoparticles. Due to the small diameter, the melting point of these nanoparticles is far below the melting point of the bulk metal[38-40]. This suggests that the nanoparticles are in a liquid state during the decomposition and they can easily change their shape. In particular, the small metal nanoparticles with diameters smaller than the inner tube diameter can diffuse into the cavities due to nanocapillarity[22,41]. Constraining forces from the walls of the encapsulating nanotube would confine the catalyst particle to a fixed diameter; this process is shown in **Figure 7**. The diameter restriction provided by the inner nanotube walls also explains why the widths of all the particles observed in the middle of nanotubes matched the inner diameter of those CNTs. When the carbon concentration inside the metallic particle exceeds supersaturation, segregation of graphite layers takes place leading to CNT growth.

## 4. Conclusions

The effect of pre heating of catalysts on carbon nanotube growth has been studied. The study reveals that the pre heating strongly affects not only degree of graphitization but also that the nanotube growth density depends on the catalyst heat treatment before the synthesis. The CNTs grown on Fe catalyst pre heated at 900°C reveals the best result in terms of growth density as well as the degree of graphitization. Heat treatment at 1000°C results in low growth density and the CNTs grown after pre heating at 850°C exhibit the worst degree of graphitization. CNT growth with Fe catalyst occurs primarily by the tip-growth mechanism and HRTEM studies prove that the internal structures of the grown materials are multiwall carbon nanotubes with partial catalyst filling.

**Acknowledgement**

We are grateful to Dr. B. Mishra from the department of Geology & Geophysics, IIT Kharagpur for his help with the Raman measurement. J. Sengupta is thankful to CSIR for providing the fellowship.

**Pre-heating effect on the catalytic growth of partially filled carbon nanotubes by chemical vapor deposition**

Joydip Sengupta and Chacko Jacob

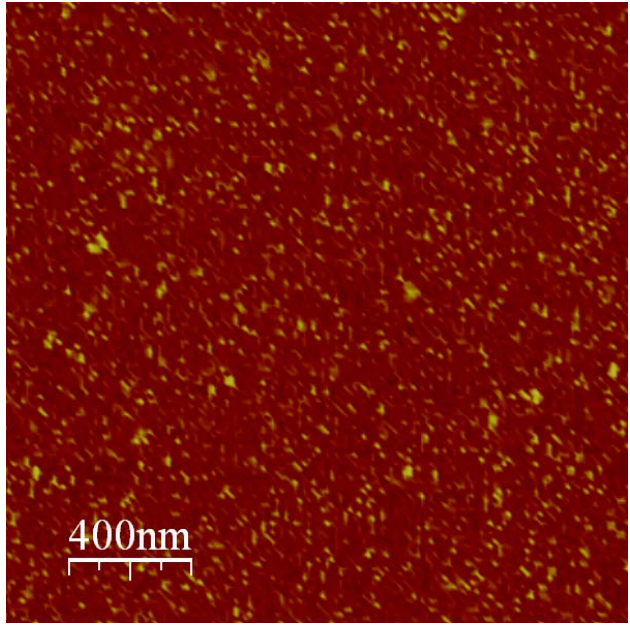

**Figure 1a** An AFM image of the as deposited Fe film on Si (111) substrate.

**Pre-heating effect on the catalytic growth of partially filled carbon nanotubes by chemical vapor deposition**

Joydip Sengupta and Chacko Jacob

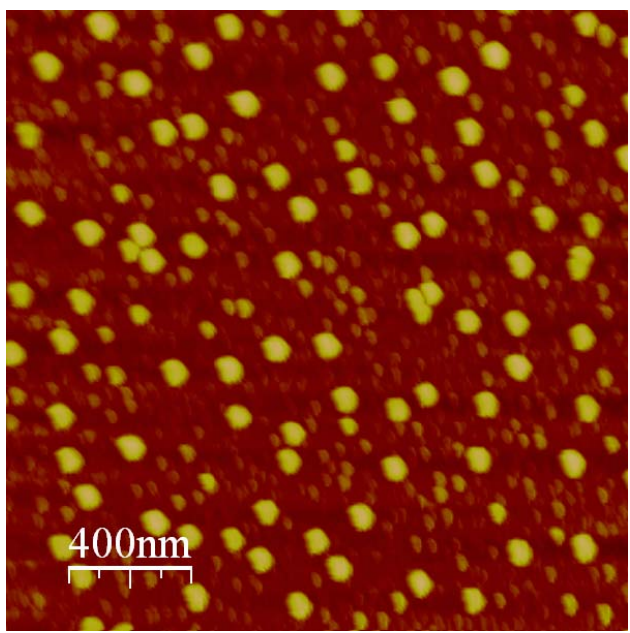

**Figure 1b** An AFM image of Fe islands formed upon pre-heating the substrate at 850°C.

**Pre-heating effect on the catalytic growth of partially filled carbon nanotubes by chemical vapor deposition**

Joydip Sengupta and Chacko Jacob

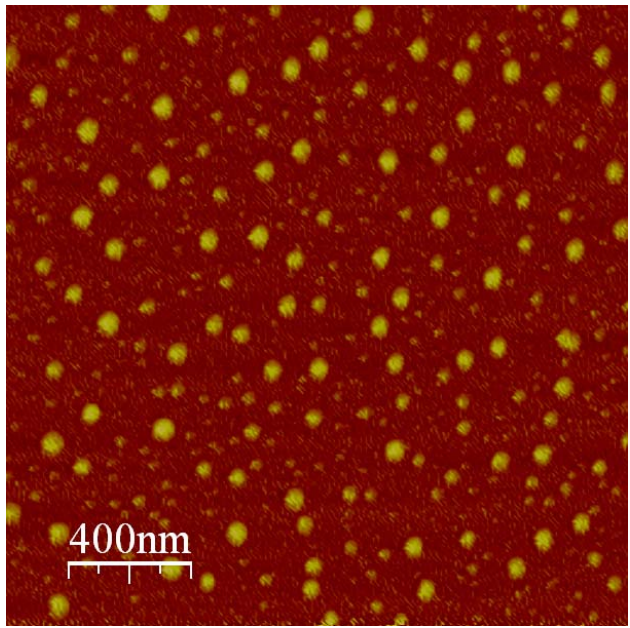

**Figure 1c** An AFM image of Fe islands formed upon pre-heating the substrate at 900°C.

**Pre-heating effect on the catalytic growth of partially filled carbon nanotubes by chemical vapor deposition**

Joydip Sengupta and Chacko Jacob

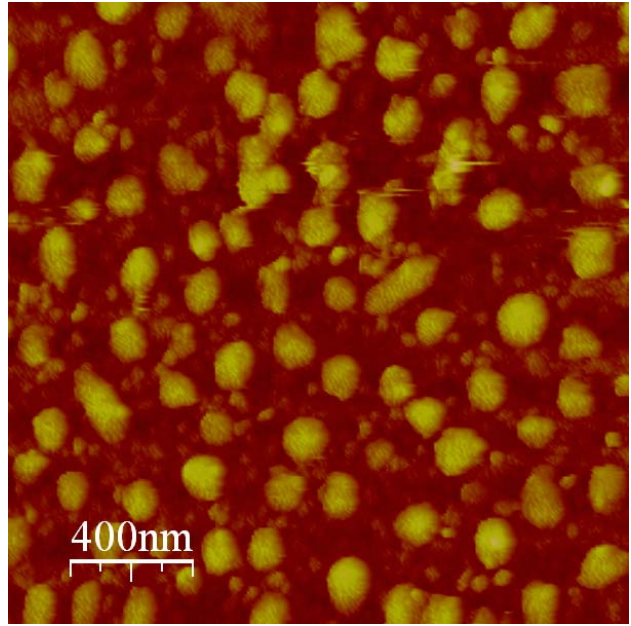

**Figure 1d** An AFM image of Fe islands formed upon pre-heating the substrate at 1000°C.

**Pre-heating effect on the catalytic growth of partially filled carbon nanotubes by chemical vapor deposition**

Joydip Sengupta and Chacko Jacob

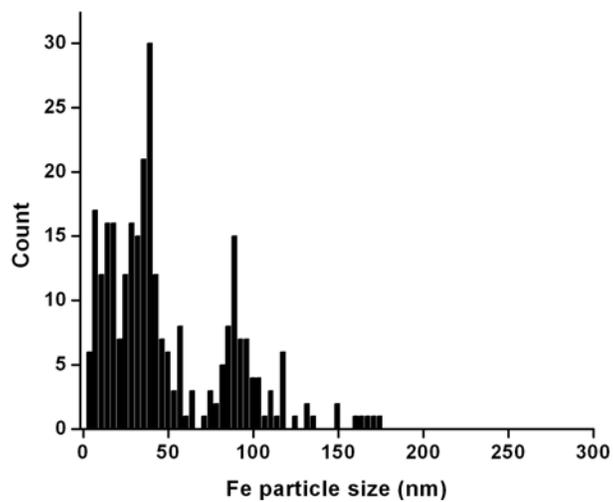

**Figure 2a** The distribution histogram of the Fe particles after pre-heating at 850°C.

**Pre-heating effect on the catalytic growth of partially filled carbon nanotubes by chemical vapor deposition**

Joydip Sengupta and Chacko Jacob

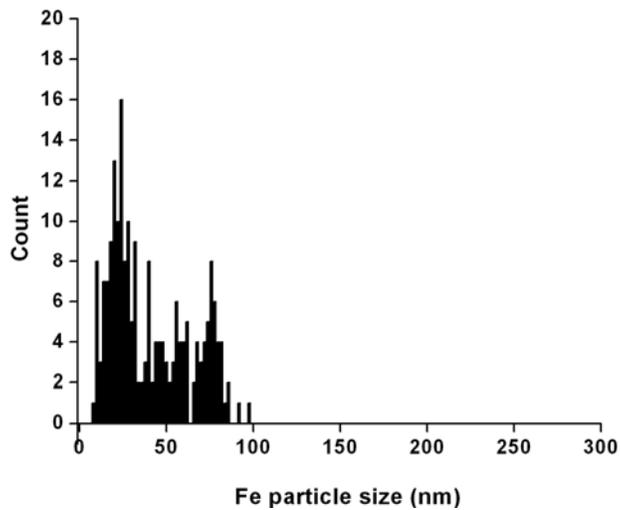

**Figure 2b** The distribution histogram of the Fe particles after pre-heating at 900°C.

**Pre-heating effect on the catalytic growth of partially filled carbon nanotubes by chemical vapor deposition**

Joydip Sengupta and Chacko Jacob

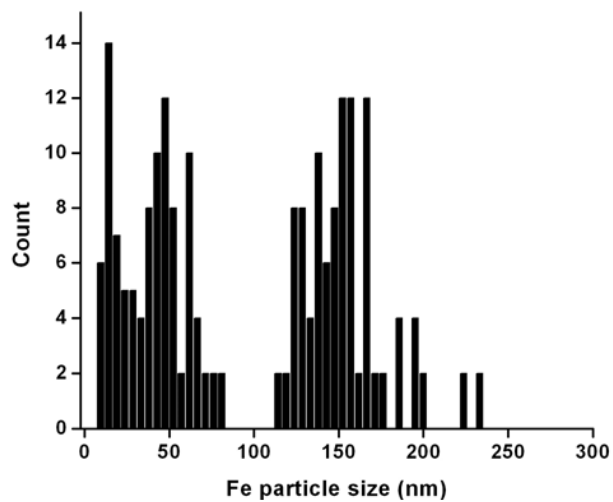

**Figure 2c** The distribution histogram of the Fe particles after pre-heating at 1000°C.

**Pre-heating effect on the catalytic growth of partially filled carbon nanotubes by chemical vapor deposition**

Joydip Sengupta and Chacko Jacob

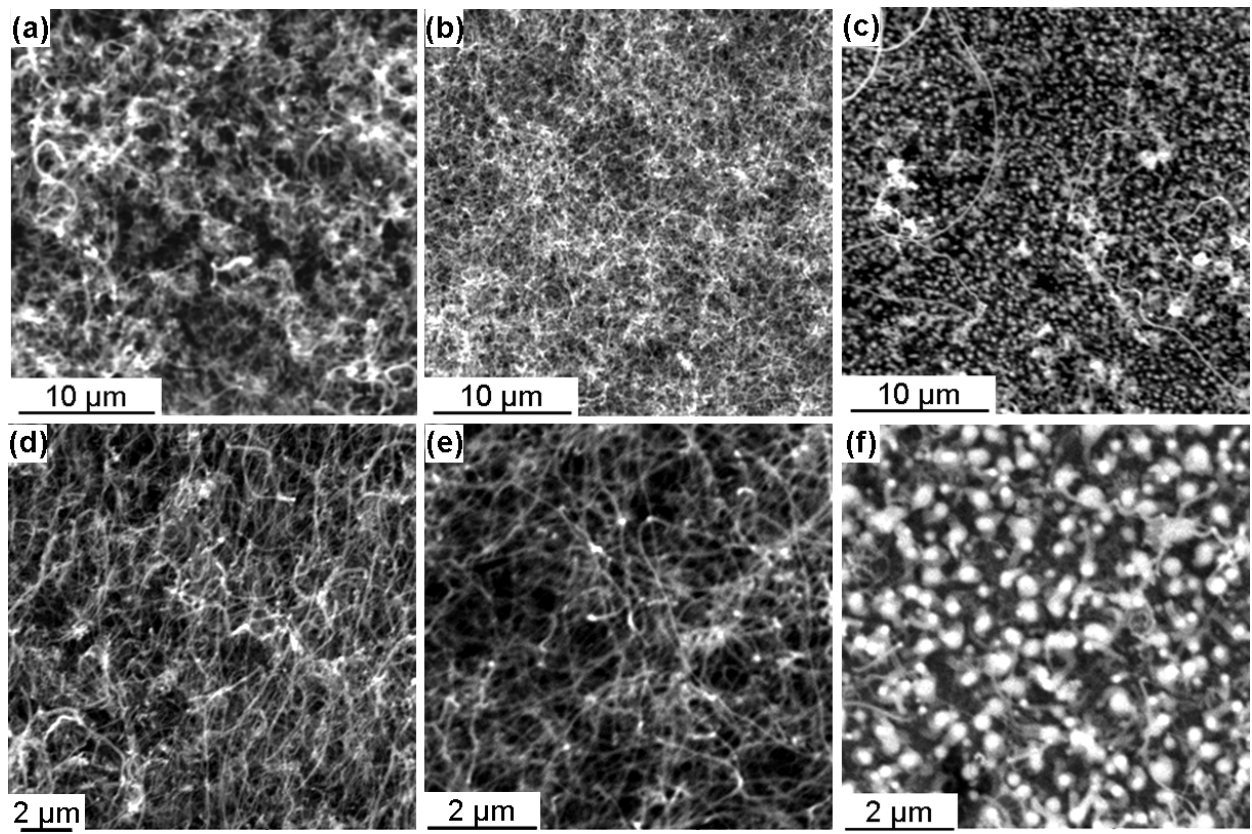

**Figure 3** SEM images of carbon nanotube arrays grown from Fe films after pre-heating in hydrogen at different temperatures. (a) and (d) 850°C; (b) and (e) 900°C; (c) and (f)1000°C.

**Pre-heating effect on the catalytic growth of partially filled carbon nanotubes by chemical vapor deposition**

Joydip Sengupta and Chacko Jacob

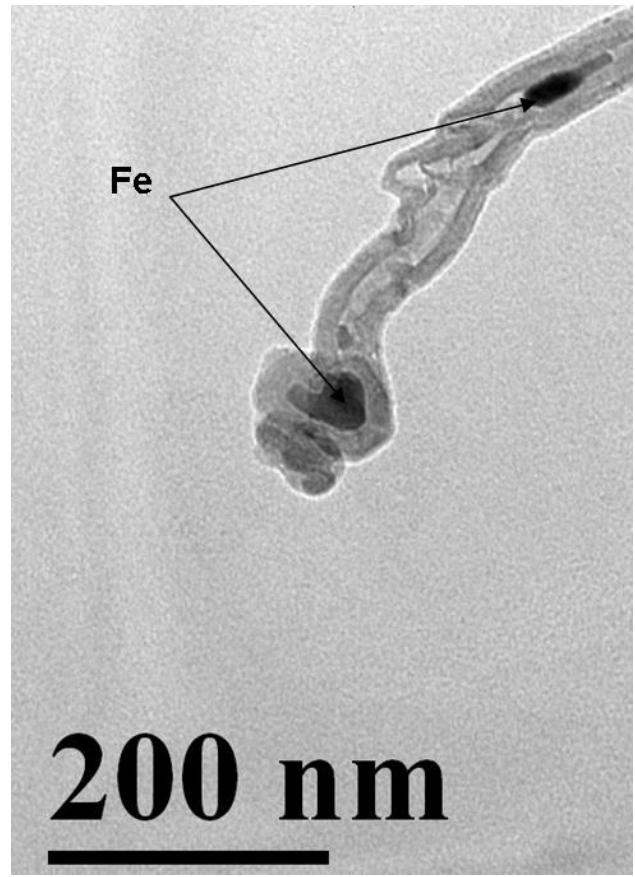

**Figure 4a** HRTEM image of the iron encapsulated CNT grown by the APCVD method using Fe catalyst after pre-heating at 850°C.

**Pre-heating effect on the catalytic growth of partially filled carbon nanotubes by chemical vapor deposition**

Joydip Sengupta and Chacko Jacob

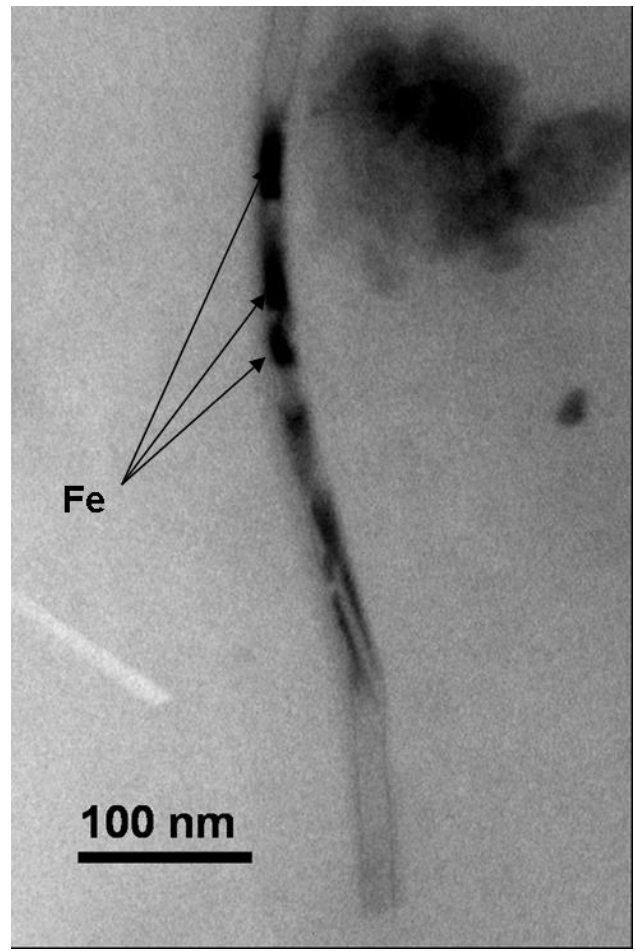

**Figure 4b** HRTEM image of the iron encapsulated CNT grown by the APCVD method using Fe catalyst after pre-heating at 900°C.

**Pre-heating effect on the catalytic growth of partially filled carbon nanotubes by chemical vapor deposition**

Joydip Sengupta and Chacko Jacob

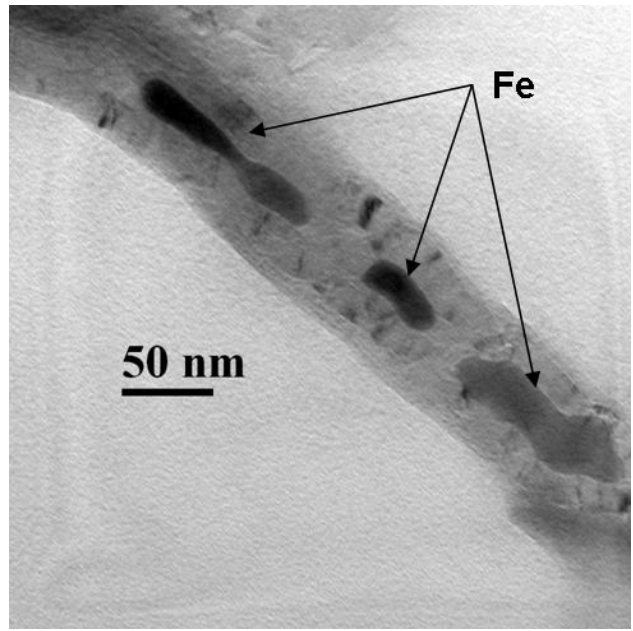

**Figure 4c** HRTEM image of the iron encapsulated CNT grown by the APCVD method using Fe catalyst after pre-heating at 1000°C.

**Pre-heating effect on the catalytic growth of partially filled carbon nanotubes by chemical vapor deposition**

Joydip Sengupta and Chacko Jacob

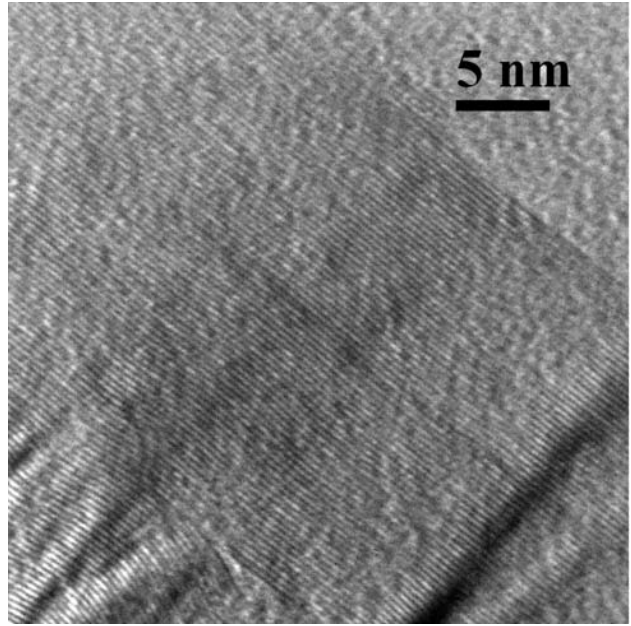

**Figure 4d** HRTEM lattice image of the CNT grown by the APCVD method using Fe catalyst after pre-heating at 850°C.

**Pre-heating effect on the catalytic growth of partially filled carbon nanotubes by chemical vapor deposition**

Joydip Sengupta and Chacko Jacob

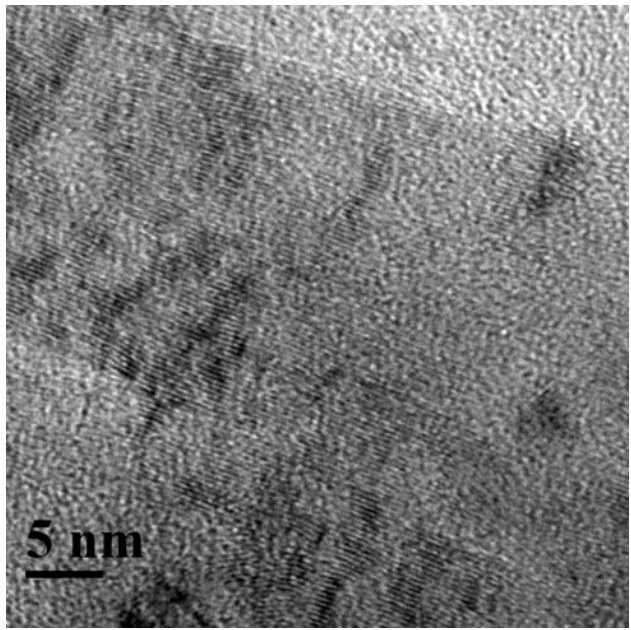

**Figure 4e** HRTEM lattice image of the CNT grown by the APCVD method using Fe catalyst after pre-heating at 900°C.

**Pre-heating effect on the catalytic growth of partially filled carbon nanotubes by chemical vapor deposition**

Joydip Sengupta and Chacko Jacob

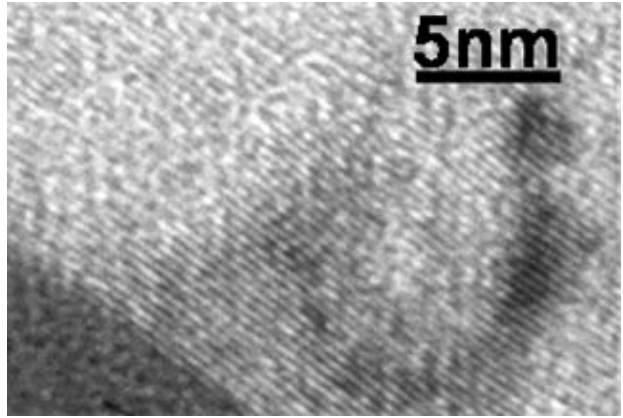

**Figure 4f** HRTEM lattice image of the CNT grown by the APCVD method using Fe catalyst after pre-heating at 1000°C.

**Pre-heating effect on the catalytic growth of partially filled carbon nanotubes by chemical vapor deposition**

Joydip Sengupta and Chacko Jacob

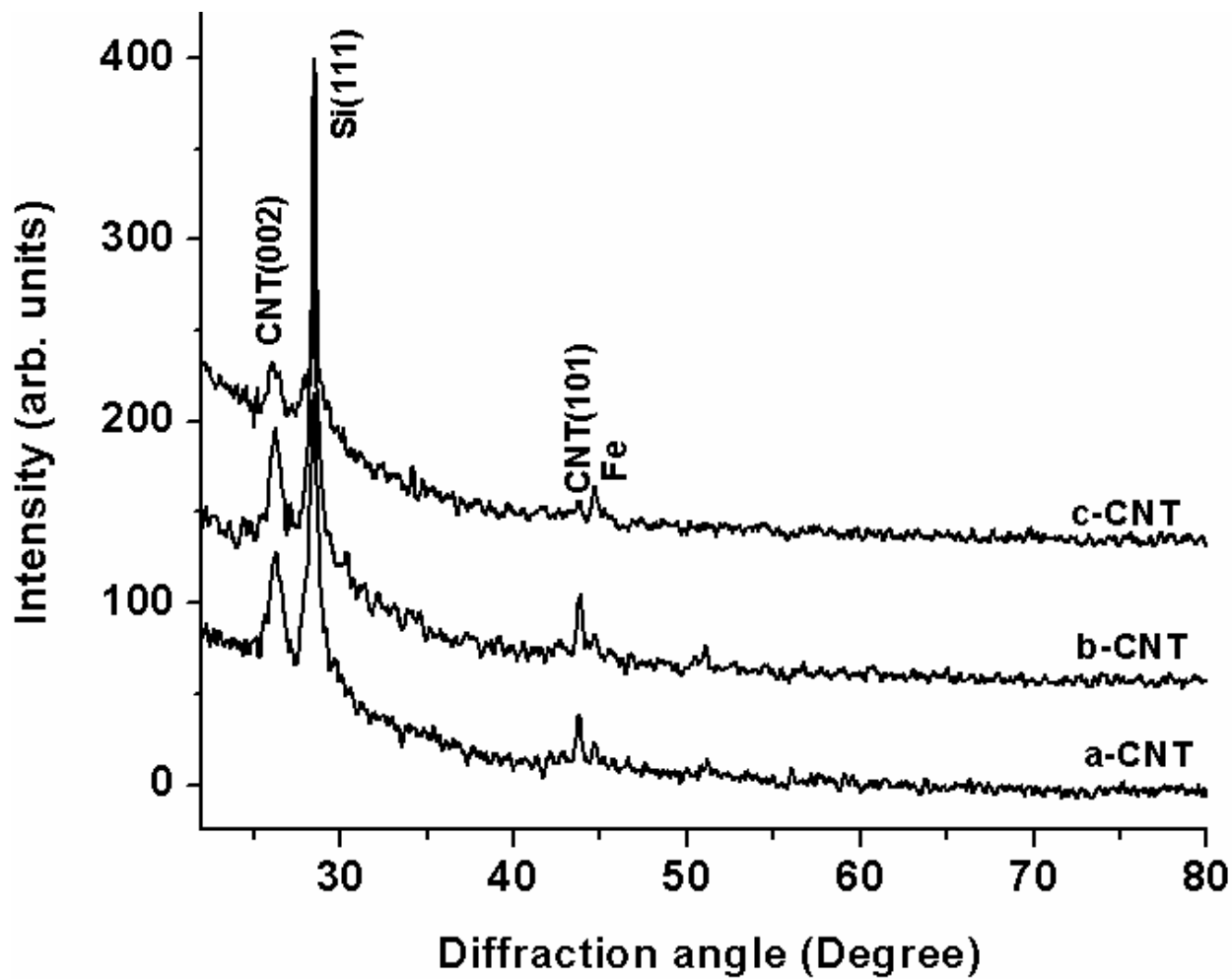

**Figure 5** X-ray diffraction spectra of MWNTs grown on Si (111) substrate using Fe catalyst after pre-heating at different temperatures: (a-CNT) 850°C; (b-CNT) 900°C; (c-CNT) 1000°C.

**Pre-heating effect on the catalytic growth of partially filled carbon nanotubes by chemical vapor deposition**

Joydip Sengupta and Chacko Jacob

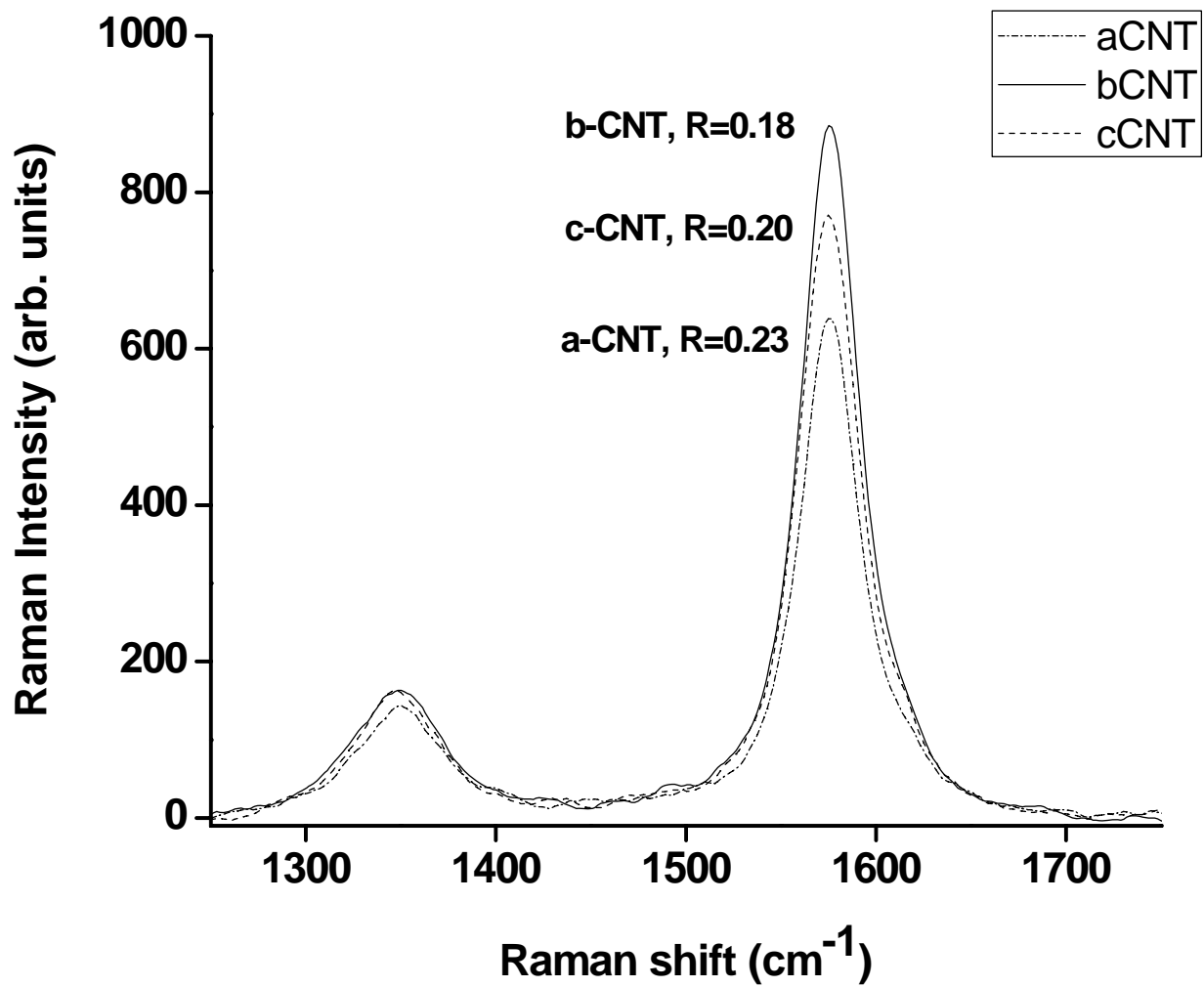

**Figure 6** Raman spectra (514.5 nm excitation) of MWNTs after pre-heating at different temperatures: (a-CNT) 850°C; (b-CNT) 900°C; (c-CNT) 1000°C.

**Pre-heating effect on the catalytic growth of partially filled carbon nanotubes by chemical vapor deposition**

Joydip Sengupta and Chacko Jacob

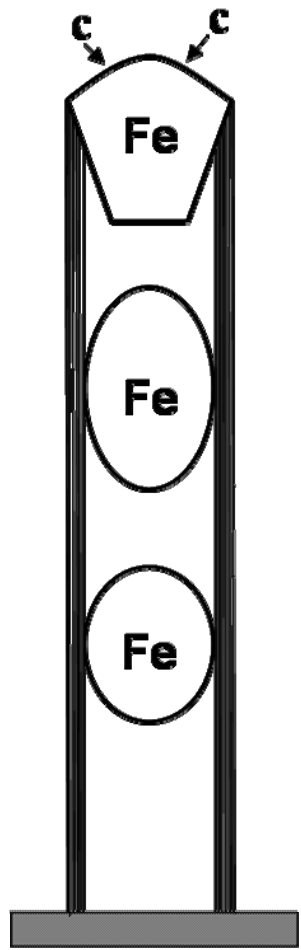

**Figure 7** The proposed growth model for the formation of partially catalyst filled MWNTs by APCVD using Fe as a catalyst.